\documentclass[apj]{emulateapj}
\begin{document}
\bibliographystyle{apj}

\title{Three-dimensional Spontaneous Magnetic Reconnection}

\author{Andrey Beresnyak}
\affil{Naval Research Laboratory, Washington, DC 20375}

\begin{abstract}
Magnetic reconnection is best known from observations of the Sun where it causes
solar flares \citep{sweet1969,parker1957,dungey1961}. Observations estimate the reconnection rate a small, but non-negligible fraction of the Alfv\'en speed, so-called fast reconnection. Until recently, the prevailing pictures of reconnection were referring to either resistivity or plasma microscopic effects, which was contradictory
to the observed rates. The alternative picture was either reconnection due to the stochasticity of magnetic field lines in turbulence \citep{lazarian1999} or the tearing instability of the thin current sheet \citep{biskamp1986,loureiro2007}. In this paper I simulated long-term three-dimensional nonlinear evolution of a thin, planar current sheet subject to fast oblique tearing instability using direct numerical simulations
of resistive-viscous MHD. The late-time evolution resembles generic turbulence with -5/3 power spectrum and scale-dependent anisotropy, so I conclude that the tearing-driven reconnection becomes turbulent reconnection. 
The turbulence is local in scale, so microscopic diffusivity should not affect large-scale
quantities. This is confirmed by convergence of the reconnection rate towards $\sim 0.015 v_A$ with increasing Lundquist number. In this spontaneous reconnection with mean field and without driving the dissipation rate per unit area also converge to $\sim 0.006 \rho v_A^3$, the dimensionless constants
$0.015$ and $0.006$ are governed only by self-driven nonlinear dynamics of the sheared magnetic field. Remarkably, this also means that thin current sheet has a universal fluid resistance depending only on its length to width ratio and to $v_A/c$.
\end{abstract}

\maketitle

\section{Introduction} 

\begin{figure}[t]
\begin{center}
\includegraphics[width=0.6\columnwidth]{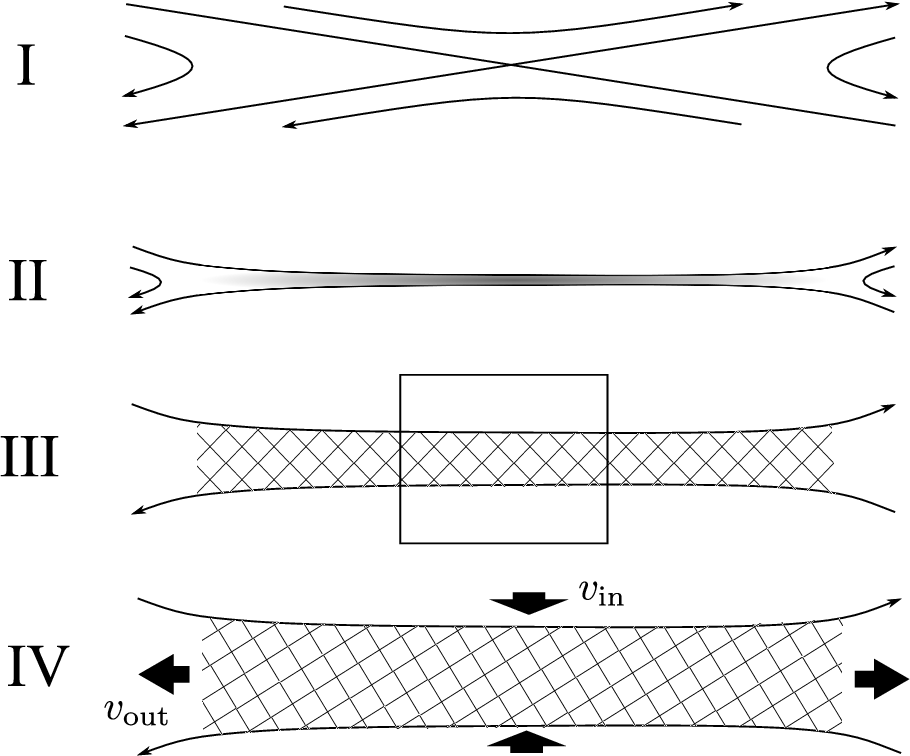}
\end{center}
\caption{A cartoon of high-Lundquist number magnetic reconnection. Magnetic X-point (I) collapses into a thin current sheet, (II), which goes unstable and produces turbulent current layer (III), expanding with reconnection rate $v_r$, until it develops an outflow and reach quasi-stationary state (IV).
Our paper discusses (III), we simulate a zoom-in of the current layer in a box which initially
appear as a planar current layer.}
\label{layer}
\end{figure}

Current sheets are abundant in magnetized plasmas. Similar to thin vortices of hydrodynamics, they are naturally
created by the nonlinear evolution of the conductive fluid \citep{parker1994,biskamp2000,priest2000}. Magnetic X-points 
naturally evolve into current sheets due to currents mutual attraction, creating the so-called Y-point configuration (Fig.~\ref{layer}). Perhaps one of the most conspicuous
phenomenon associated with current sheets in plasmas are solar flares, the bursts of radiation of up to
$6\times 10^{33}$ ergs in X-rays. Following the big solar flare, the coronal mass ejections (CME) occurs, hinting
to the global rearrangement of the magnetic field, which is called magnetic reconnection. Another well-known process is a magnetospheric storm, which is a perturbation of magnetosphere, associated with reconnection in the magnetotail. While CME demonstrates
that there was a topological rearrangement of the magnetic field, the careful observations near the flare site
typically estimate the rate of inflow of magnetic field lines, called reconnection rate, to a $0.001-0.1$ fraction
of the Alfv\'en speed $v_A=B/\sqrt{4\pi \rho}$ \citep[see, e.g.,][]{dere1996}.

\begin{figure}
\begin{center}
\includegraphics[width=0.7\columnwidth]{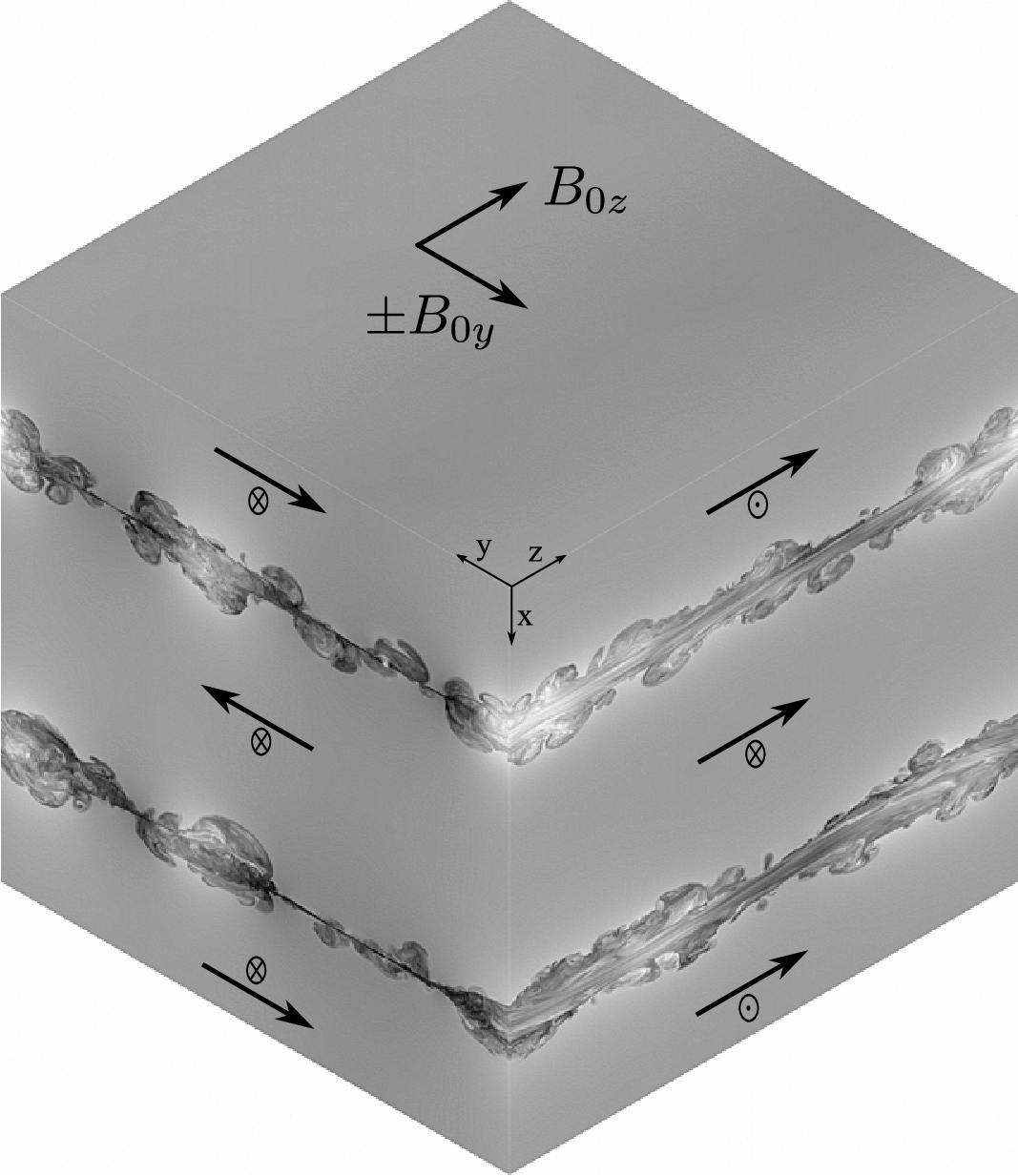}
\end{center}
\caption{This figure demonstrates the all-periodic box reconnection setup with two current layers,
reconnecting field $\pm B_{y0}$ and the imposed mean field $B_{z0}$, the magnitude of $B$ is shown as grayscale on the surface of the box.}
\label{cube}
\end{figure}
In a well-conductive plasma, one might expect that current sheets are non-dissipative and, therefore, invisible.
Indeed, the Sweet-Parker (SP) model \citep{sweet1969,parker1957} predicts very low reconnection
rate for most astrophysical and space magnetic configurations. The dimensionless number characterizing plasma conductivity is the Lundquist number $S=L v_A/\eta$, where $L$ is the length of the layer $\eta$ is magnetic diffusivity. High Lundquist number
means the magnetic resistive decay time, $L^2/\eta$, is much larger than Alfv\'en crossing time, $L/v_A$. For
laminar thin current sheets, the Sweet-Parker (SP) model \citep{sweet1969,parker1957} predicts reconnection rate
of $v_A/\sqrt{S}$, this enhancement compared to resistive diffusion is due to the fact that magnetic field
diffuses only through a thin width of the current sheet $L/\sqrt{S}$. This speed, however, is extremely low
for most astrophysical and space magnetic configurations. The SP model, in the limit of very high $S$, becomes
consistent with the so-called frozen-in condition of the ideally conductive fluids. The SP prediction, therefore, contradicts the idea that the discontinuity in the magnetic field may result in an arbitrary
reconnection rate independent on resistivity, as in the Syrovatskii's model \citep{Syrovatskii1971}. The search for fast
reconnection have shifted towards microscopic effects beyond MHD, e.g. effects in collisionless plasmas
\citep{drake2006,daughton2011,che2011}. Alternative approaches suggested that in presence of turbulence
the magnetic field lines will be stochastic \citep{lazarian1999,kowal2009,eyink2011b,eyink2013} which would lead to fast reconnection, which have implications for particle acceleration \citep{LB09b,BL15,BH16}. The study of the resistive tearing instability in the thin current sheet \citep{biskamp1986}, surprisingly resulted in a conclusion that it becomes faster and not slower with decreasing resistivity \citep{loureiro2007,huang2010,loureiro2012} at sufficiently high $S$. It has become clear that SP model
is problematic at high $S$ because thin SP current sheets are unstable above the critical Lundquist number
$S=L v_A/\eta \sim 10^4$. In a two-dimensional (2D) resistive reconnection scenario, secondary instability
of the current sheet between magnetic islands will result in a resistivity-independent reconnection \citep{uzdensky2010}. The two-dimensional (2D) MHD simulations
measured reconnection speeds around $0.01\div 0.03v_A$ \citep{loureiro2012, huang2010} and observed hierarchical formation and ejection
of plasmoids. Plasma simulations demonstrated that
collisionless thin current layers are also unstable \citep{daughton2009}. 

In this paper I did three-dimensional (3D) simulations of thin current sheet with significant imposed mean field, starting with oblique tearing and developing into nonlinear phase which I called spontaneous turbulent reconnection. 

In what follows Section 2 describes simulation setup, Section 3 overview the results of simulations
in terms of bulk average quantities, such as total energy budget and its evolution, Sections 4 and 5 describes local properties of turbulence, spectrum and anisotropy. Section 6 comments on the global nature of the perturbation of the magnitude of the magnetic field (slow mode). Section 7 discusses
in some detail the differences in reconnection picture in 2D and 3D cases. Section 8 proposes
phenomenological model for the reconnection rate. Section 9 discusses implications for particle acceleration, Section 10 points out that thin current layer could be viewed from electromagnetic
viewpoint as having non-zero resistance per unit length, even in the limit of vanishing resistivity.
Section 10 is a Discussion.


\section{Problem setup}

One of the simplest setups to study nonlinear development of tearing is a periodic setup with the mean field $B_{z0}$ threading the box, reconnecting field $\pm B_{y0}$ changing sign in the $x$ direction. I also consider incompressible case, in which situation the problem has only two defining dimensionless numbers: 1) the Lundquist number $S$ 2) the ratio $B_{y0}/B_{z0}$. 
\begin{table}[t]
\begin{center}
\caption{Spontaneous reconnection MHD experiments}
  \begin{tabular*}{1.00\columnwidth}{@{\extracolsep{\fill}}c c c c c c}
    \hline\hline
Run  & $N^3$ & Dissipation & $S$ or $S_4^{0.4}$ & $B_{y0}/B_{z0}$ & $v_r/v_A$\\
   \hline
N1   & $576^3$ & $-3.6\cdot10^{-4}k^2$ & $1.7\times 10^4$ &  1.0 & 0.0124\\
\hline
H1B1 & $576^3$ & $-2.4\cdot10^{-9}k^4$ & $2.5\times 10^4$ &  0.5 & 0.0214 \\
H1B2 & $576^3$ & $-2.4\cdot10^{-9}k^4$    & $2.5\times 10^4$ &  1.0 & 0.0210\\
H1B3 & $576^3$ & $-2.4\cdot10^{-9}k^4$ & $2.5\times 10^4$ &  2.0 & 0.0187 \\
\hline
N2 & $768^3$ & $-2.5\cdot10^{-4}k^2$   & $2.5\times 10^4$ & 1.0 & 0.0117\\
H2 & $768^3$ & $-9.4\cdot10^{-10}k^4$   & $3.7\times 10^4$ & 1.0 & 0.0183\\
\hline
N3 & $1152^3$ & $-1.4\cdot10^{-4}k^2$   & $4.4\times 10^4$ & 1.0 & 0.0155\\
H3 & $1152^3$ & $-9.7\cdot10^{-10}k^4$   & $3.6\times 10^4$ & 1.0 & 0.0146\\
\hline
N4 & $1536^3$ & $-9.8\cdot10^{-5}k^2$   & $6.4\times 10^4$ & 1.0 & 0.0154\\
H4 & $1536^3$ & $-3.7\cdot10^{-10}k^4$   & $5.4\times 10^4$ & 1.0 & 0.0144\\

   \hline
\end{tabular*}
  \label{experiments}
\end{center}
\end{table}

This very simple geometry physically corresponds to the initial (albeit already nonlinear) development of
tearing before the outflow becomes important. The cartoon on Fig.~\ref{layer} illustrates the typical progression of high-$S$
spontaneous reconnection by showing a cut perpendicular to the current and global mean field direction: magnetic configuration with the X-point (I) may develop into the thin current sheet (II), the latter develops
instability, the instability goes nonlinear and produces turbulent current layer (III) which later
expands and produces classic picture with inflow and outflow (IV). I studied the initial regime before the outflow becomes important, designated as regime III on Fig.~\ref{layer}, which is especially 
interesting because it has highest volumetric dissipation rate.

I used all-periodic setups, so I actually simulated two current layers, and all the results
is the average between the properties of these two layers. Fig.~\ref{cube} demonstrates the setup and shows the simulation snapshot during the development of the nonlinear phase, with the magnitude of $B$ is shown by grayscale on the surface of the box.
One of the numerical challenges in studying 3D spontaneous reconnection was to break through the critical Lundquist number barrier of $10^4$, which require sufficiently big boxes. 
In simulations with imposed large-scale perturbation, such as \citet{daughton2011,Oishi2015,Huang2016}
the Lundquist number is directly estimated using the size of the perturbation, which is typically the
box size. In these cases simulations try to reproduce the whole current layer in a Sweet-Parker
configuration in regime IV of Fig.~\ref{layer}. I, on the other hand, try to simulate a zoom-in of
the middle of the current layer in regime III which initially look like a planar current sheet
but at later times will develop large scale perturbations. If I define Lundquist number using the box size, as $S=v_{Ay} L/\eta$, and the system size $L_S$ is actually bigger than the zoom-in box size, the
Lundquist number of the whole system $S_S=v_{Ay} L_S/\eta$ is larger than S which I quote in Table 1. I can, therefore,
safely assume that the larger system is unstable to tearing, just as my planar current
sheet is unstable. A subtle difference of these two types of setup is that the simulations with global
large-scale initial perturbations aim to describe stationary regime IV at later times, $t\gg L_S/V_A$,
with a finite, albeit large, S, while my planar current sheet setup aim
to simulate earlier times, $t<L_S/V_A$, when the global outflow did not have time to form yet.
I also assume that $L_S \gg L$ i.e. the global Lundquist number $S_S$ is asymptotically large, so
I can ignore gradients from the large-scale setup of the system and only impose small-scale perturbations. The end time of my simulations was determined by the development
of large-scale structures, as long as these structures become comparable with the box size, the box
size deemed not sufficient, similarly to simulations of nonlinear Kelvin-Helmholtz instability. 
  
I used pseudospectral code with explicit dissipation coefficients -- viscosity and magnetic diffusivity,
equal to each other. The code solves incompressible resistive-viscous MHD equations, does not have
inherent dissipation or dispersion grid errors and is mostly described in \citep{B14a}, except in
the present paper I used full MHD, not reduced MHD. These simulations are DNS, e.g. they are well-resolved with the dimensionless maximum wavenumber satisfying $k_{\rm max}l_\eta > 1$, where $l_\eta=(\eta^3/\epsilon)^{1/4}$ is the dissipation (Kolmogorov) scale\footnote{Moderately under-resolved cases exhibit visible ringing at grid scale, which happened in simulation H2 (Table~1), this simulation wasn't included in the reconnection rate measurements.}.

Normal viscosity and magnetic diffusivity $\eta\nabla^2 B$ was used in one series of simulations and the hyper-diffusivity of the forth order,
$\eta_4 \nabla^4 B$ in the other. The Lundquist number was defined in terms of the box size $L=2\pi$ and the reconnecting field $v_{Ay}=1$ as $S=v_{Ay} L/\eta$ and the hyper-Lundquist number as $S_4=v_{Ay} L^3/\eta_4$. The Lundquist number $S$ and the hyper Lundquist number $S_4$ that give the same dynamical range of scales between dissipation scale and outer scale, e.g. the ratio $L/l_\eta$, are related by $S=S_4^{4/10}$, assuming Kolmogorov scaling. The rationale behind the two dissipation schemes
was to check the influence not only on $S$, but also on the dissipation functional form to the bulk quantities, such as reconnection and dissipation rate. The magnetic Prandtl number was unity: $Pr_m=\nu/\eta=1$, which was motivated by the desire to reach the highest possible $S$ while staying well-resolved. As I will show below, the amount of magnetic and kinetic dissipation were close to each other, which is typical for an ordinary MHD cascade. 
\begin{figure}[t]
\begin{center}
\includegraphics[width=1.0\columnwidth]{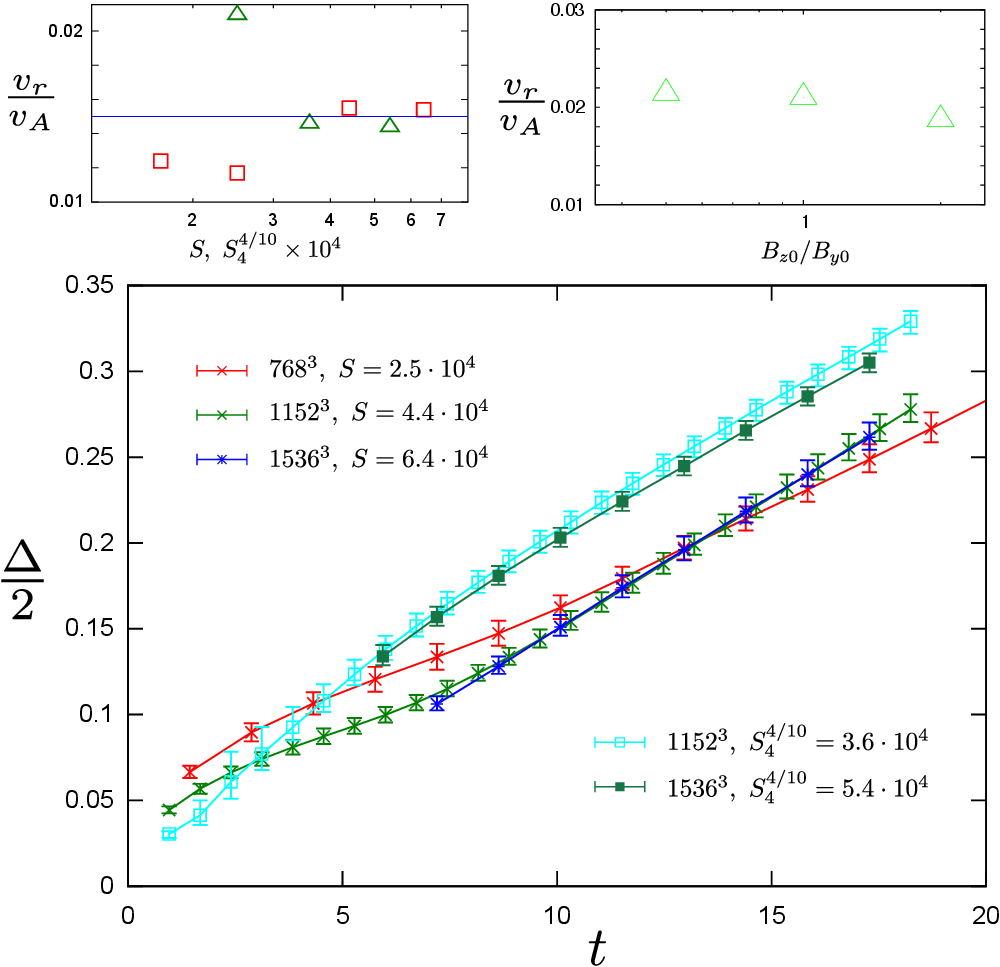}
\end{center}
\caption{The time evolution of the current layer width $\Delta$ (bottom) and the inferred reconnection speed
as a function $S$ (top left) and the ratio of $B_{z0}/B_{y0}$ (top right). The error bars were obtained by varying the current density threshold by a factor of two.
The difference in the initial evolution of hyper- and normal-diffusion cases is due to much faster
development of hyper-resistive oblique tearing compared to normal oblique tearing.
The variation of $v_r$ around $\sim$13\% with $B_{z0}/B_{y0}$ varying by a factor of 4 and was within measurement error.}
\label{width}
\end{figure}

The simulations were set up with a thin current sheet with Harris profile and seeded with small initial
perturbations, $\sim 10^{-6}$ of the magnetic energy. These perturbations subsequently evolved
due to oblique tearing near the current sheet itself, the bulk of the volume was almost undisturbed.
At later times this evolved into fully nonlinear turbulent current layer.
The boundary between the undisturbed volume, which had nearly zero current and the turbulent current layer was fairly well-defined, as shown on Fig.~\ref{cube}.
The inside of the current layer was determined as opposite to the undisturbed fluid where
current density is always small. The point at which the current density exceeds a certain threshold
in magnitude was regarded as the beginning of the current layer. I varied current density thresholds
to study the dependence on the measurement of the layer width $\Delta$. The procedure to obtain
the errors for that measurement was to vary the lower and upper thresholds for current by a factor of two.
The list of all simulations is presented on Table~1.

\section{Evolution}

The evolution of the current layer width and the inferred reconnection rate are shown on Fig.~\ref{width}. 
The system initially contained the energy density of the opposing field $B_{y0}^2/8\pi$, which was free to dissipate and the energy density of the mean imposed field $B_{z0}^2/8\pi$, which had to conserve due to conservation of total flux through x-y plane.
After subtracting the latter contribution, a I designate a dimensionless free energy density as 
\begin{equation}
w=(4\pi \rho v^2+B^2-B_{z0}^2)/B_{y0}^2,
\end{equation}
which is unity in the undisturbed fluid. After $t\approx 2$ turbulence in the layer fully develops, and the average $w$ within the layer, $w_t \approx 0.6$, while the undisturbed part still have $w=1$.
The dissipation of $w_d\equiv 1-w_t=0.4$ fraction of energy happens during development of turbulence and stays approximately constant.

I inferred reconnection rate as the growth speed of the current layer width $v_r=d \Delta/dt$. This
is different from a conventional definition as an inflow speed in stage IV (Fig.~1). In stage III,
however it is a meaningful definition, in terms of how much free magnetic energy is available to the
system per unit time per unit area of the current layer. From this energetic viewpoint inflow definition
and my definition are similar.

$V_r$ was around $0.015 v_{Ay}$ for high Lundquist numbers and appear to be only weakly dependent on the imposed mean field $B_{z0}$ (Fig.~\ref{width}). The dissipation rate per unit area from both sides of the current sheet (note a factor of two) can be calculated from $w_d$ and $v_r$ as 
\begin{equation}
\epsilon_S=2 w_d v_r (1/2) \rho v_{Ay}^2 \approx 0.006 \rho v_{Ay}^3,
\end{equation}
note that conventional dissipation rate per unit mass, traditionally used in theory of incompressible turbulence will be expressed using current layer width $\Delta$
as 
\begin{equation}
\epsilon=(1/\rho)\epsilon_S/\Delta=w_d v_r v_{Ay}^2/\Delta,
\label{eps_form}
\end{equation}
and will depend on time. The turbulence in the expanding current layer is not stationary turbulence in a sense that it grows in volumes and produces turbulent energy
as well as dissipates energy. The outer scale of this turbulence also grows in time.

\begin{figure}[t]
\begin{center}
\includegraphics[width=0.9\columnwidth]{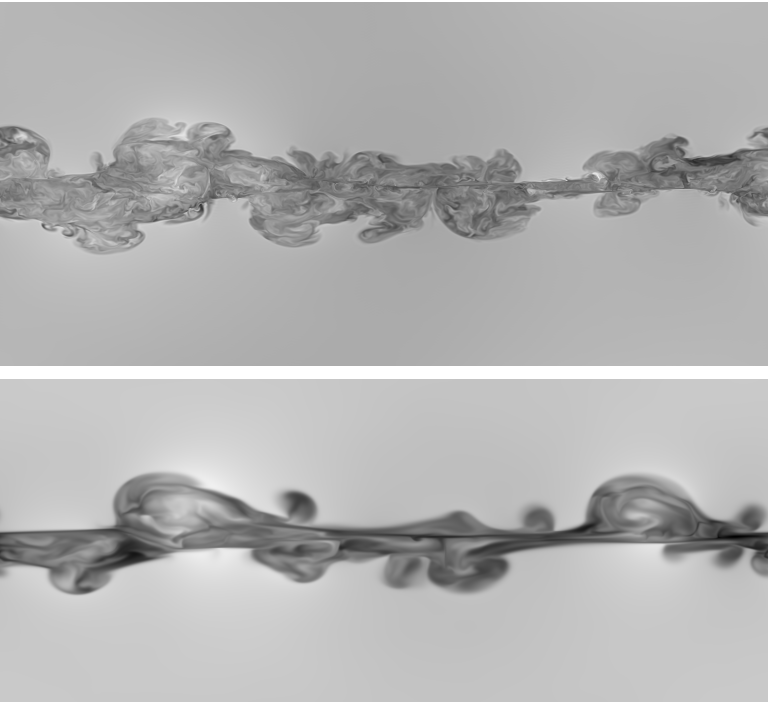}
\end{center}
\caption{Zoom-in x-z slices of the turbulent current layer. Mean magnetic field is out of the plane. Upper part is the magnitude of B in hyper-viscous simulation, lower part -- in viscous simulation.}
\label{b_slice}
\end{figure}
The contribution to turbulent fraction of the energy density $w_t\approx 0.6$ was partitioned to $\sim 
0.55$ in x and y magnetic component, $\sim 0.02-0.04$ in $\delta B_z$ component and  $\sim 0.01-0.02$ in 
kinetic energy. The turbulence in the current layer was strongly anisotropic with respect to $B_{z0}$ 
direction, with wavevector predominantly perpendicular to $z$. The $B_x$ and $B_y$ components, carrying 
most of the energy, therefore, represented Alfv\'enic perturbations, while the sub-dominant  $\delta 
B_z$ was the slow-mode (pseudo-Alfv\'en) perturbation. I discuss anisotropy in more detail in 
Section~\ref{anis},
noticing that the fact that the reconnection rate depends only weakly on $B_{z0}$ is not surprising,
since the anisotropic turbulence of Alfv\'enic perturbations also known as Alfv\'enic turbulence
or reduced MHD turbulence possesses rescaling symmetry with respect to $B_{z0}$ \citep{B12b}, which
I actually confirm in Section~\ref{anis}. The domination of Alfv\'enic perturbations in reconnection
with strong mean field is extremely important as it sheds light to fluid-like behavior in plasma simulations,
e.g. \citep{daughton2009}, in spite of significant collisionless effects. The explanation for this is
that reduced MHD is well-applicable to collisionless plasmas on scales above ion Larmor radius $r_L$,
and that plasma does not require significant collisional terms to behave like reduced MHD fluid \citep{schekochihin2009}.

\section{Spectra}
\begin{figure}[t]
\begin{center}
\includegraphics[width=1.0\columnwidth]{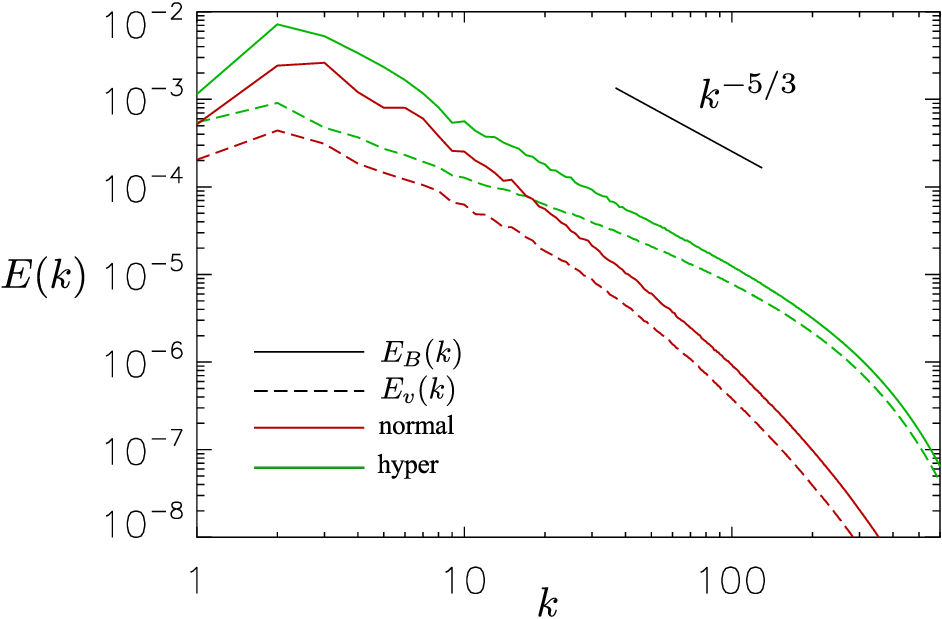}
\end{center}
\caption{The y-z power spectra of velocity and magnetic field for simulations N4 and H4. The spectral slopes were around $-1.5\div -1.7$, which is characteristic
of local-in-scale turbulence.}
\label{spectrum}
\end{figure}

The structure of the perturbed current layer looked rather turbulent (Fig.~\ref{b_slice}) with only a
small fraction of the initial current sheet structure being retained.

I defined the power spectra of turbulent perturbations in the yz plane as
\begin{equation}
E(k_l)=L^{-1} \int \hat f(\mathbf{k_l}) \hat f^*(\mathbf{k_l}) d\phi\, dx
\end{equation}
where $k_l=(k_y,k_z)$ -- a wavevector in yz plane, $\hat  f(\mathbf{k_l})$ -- Fourier transforms of either v or B.
Neglecting $k_x$ in this spectrum is necessary to get rid of the contribution from $\pm B_{y0}$ jump across the current layer which happens in the x direction. The spectrum, presented on Fig.~\ref{spectrum} has magnetic contribution dominating over kinetic on large scales, but tend to approximate equipartition on smaller scales. This is not surprising, since turbulence is driven by magnetic energy. This spectral picture, qualitatively, is characteristic for decaying magnetic turbulence, including  cases when initial field was completely random \citep[see, e.g.][]{Biskamp2003,Brandenburg2015}.

The total energy spectral slope was around $-1.5\div -1.7$, roughly consistent with Goldreich-Sridhar \citep{GS95} scaling. The slopes between -1 and -3 are indicative
of local-in-scale turbulence. Precise measurement of the spectral slope in these simulations
was difficult due to limited inertial range, however, the scale-locality would imply that at sufficiently high $S$ the inertial range scaling and anisotropy will be the same as in the homogeneous driven MHD turbulence. In the next section I test the anisotropy component
of this conjecture.

The scale-locality is a key ingredient in theories of turbulent reconnection. Indeed full
scale-locality will imply that large-scale quantities, such as reconnection rate and
dissipation rate per unit area should be independent on any microphysics. Speaking in practical
terms, if both ion Larmor radius $r_L$ and ion skip depth $d_i$ are much smaller than the minimum size of the problem -- the layer width $\Delta$, reconnection rate will
be independent on microphysics. Note that in regime (IV), stationary reconnection, $\Delta$
will become constant around $0.015L$\footnote{Assuming reconnection rate is $0.015v_A$ in regime IV, see also simulations with outflow, e.g. \citet{loureiro2012}.}. This allows us to estimate the range of applicability
of this particular mechanism of fast reconnection.
\begin{figure}[t]
\begin{center}
\includegraphics[width=1.0\columnwidth]{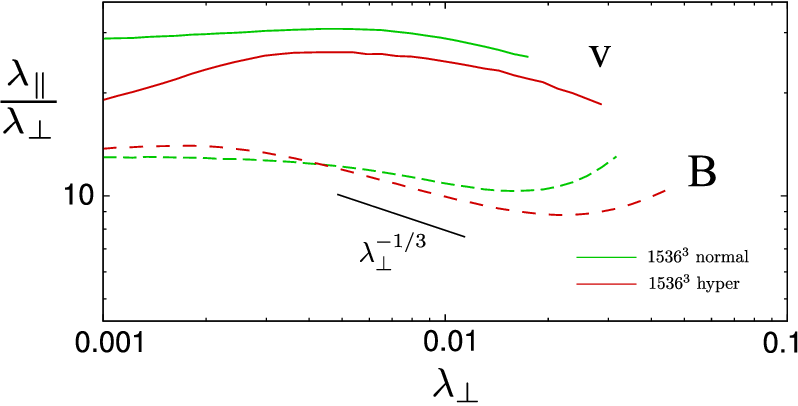}
\end{center}
\caption{Anisotropy of velocity and magnetic perturbations in the current layer measured with respect to the local field. I used conditional 2-point structure function where both points
were within the current layer. The high anisotropy, especially for the velocity field, is due to low amplitude of velocity perturbations
and approximately corresponds to the critical balance between parallel and perpendicular timescales.}
\label{anis}
\end{figure}

\section{Anisotropy}
I have calculated second-order structure functions of the turbulent $v$ and $B$ fields inside
the current layer, e.g. for velocity:
\begin{equation}
SF^2_v(l_\|,l_\perp)=\langle(v({\bf r}-{\bf l})-v({\bf r}))^2\rangle_{\bf r}.
\end{equation}
Note that I assumed that it depends only on the component of ${\bf l}$ parallel and perpendicular to the magnetic field. Two types of such measurement are possible: when
parallel direction is determined by the global mean magnetic field, in my case z direction,
or local magnetic field ${\bf B}$. Scale-dependent anisotropy of \citet{GS95} model is
observed with the local measurement \citep[see, e.g.,][]{CV00,BL09a}.

Using these structure functions for $v$ and $B$ I built correspondence between
$\lambda_\|$ and $\lambda_\perp$ by equating SF values in parallel and perpendicular direction.
More details on this type of measurement can be found in \citet{BL09a}. Fig.~\ref{anis} shows
anisotropy $\lambda_\|/\lambda_\perp$ as a function of $\lambda_\perp$. One thing to notice is that the value of anisotropy in this particular simulation, with $B_y/B_z=1$, is around 20. At the same time the RMS value of velocity perturbation is around $0.08-0.11v_{Ay}$. The interaction strength parameter $\xi=\delta v \lambda_\|/v_A \lambda_\perp$ will be around unity, i.e. these perturbations are, approximately, ``critically balanced''. This means that from MHD perspective we are dealing
with ``strong'' turbulence, i.e. nonlinear interaction terms have the same contribution as the
tension of the mean field $B_z$. Also note that this anisotropy corresponds to the angle of the field line bending $\sim 1/20$
which is much smaller than the angle of the initial stripes of developing oblique tearing ($45^o$ in the case of $B_y/B_z=1$). So the turbulence self-organizes itself into being strong and forgets
properties of the oblique tearing that initiated it.
 
The evidence for scale-dependent anisotropy is only tentative, considering rather short inertial range, the expected law of scale-dependency from \citet{GS95} is $\lambda_\|/\lambda_\perp \sim \lambda_\perp^{-1/3}$, see Fig.~\ref{anis}.

\begin{figure}[t]
\begin{center}
\includegraphics[width=1.0\columnwidth]{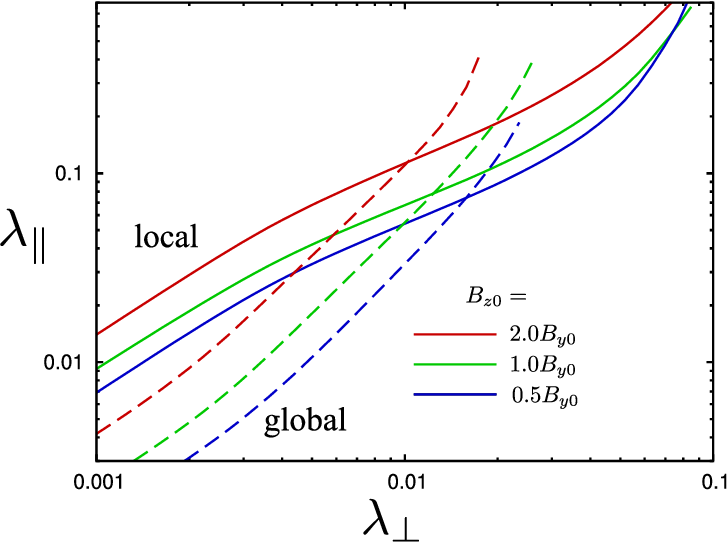}
\end{center}
\caption{Comparison of anisotropies in simulations with different imposed field. Solid lines -- measurement with respect to the local field, dashed lines
-- measurement with respect to the $z$ direction. We expect symmetry $\lambda_\|/B_{z0}=const$ in the limit of $B_{z0}>>B_{y0}$ for the measurement with respect to z direction
(RMHD symmetry). This symmetry is well satisfied between $B_{z0}=2.0 B_{y0}$ and $B_{z0}=1.0 B_{y0}$ cases, but not so well satisfied between
$B_{z0}=1.0 B_{y0}$ and $B_{z0}=0.5 B_{y0}$ due to the fact that $B_{z0}/B_{y0}$ is not asymptotically large.}
\label{anis_b}
\end{figure}

Another important indicator is how anisotropy varies with the value of the mean field $B_{z0}$,
this is presented on Fig.~\ref{anis_b}. I plotted both local and global anisotropy measurements.
Note how $\lambda_\|$ measurements have the same shape and are increasing with increasing $B_{z0}$. In a purely Alfvenic dynamics, also called reduced MHD, $\lambda_\|$ is strictly proportional to $B_{z0}$
\citep[see, e.g.,][]{B12b}. I noticed that this scaling is almost perfect between $B_{z0}=1$ and
$B_{z0}=2$ cases, which further confirms that Alfv\'enic dynamics dominates in the current layer turbulence.

\section{Diamagnetism}
\begin{figure}
\begin{center}
\includegraphics[width=0.7\columnwidth]{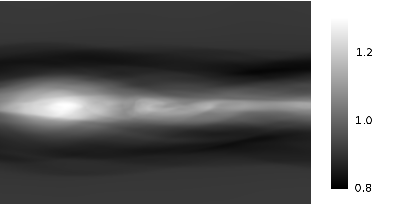}
\end{center}
\caption{Zoom in of $B_z$ averaged over $z$ in the x-y plane for simulation with $B_{z0}/B_{y0}=1$.
The width of the picture, along $y$, is $0.4 L$.
In strongly anisotropic perturbations this component represents pseudo-Alfv\'en or slow mode.}
\label{bz_slice}
\end{figure}

In the $B_{y0}/B_{z0}=1$ case, apart from Alfv\'en mode, which contained 93\% in total energy of turbulent motions, the rest 7\% are perturbations in $B_z$. These perturbations are strongly anisotropic and this component represents pseudo-Alfv\'en or slow mode.
The perturbations are energetically dominated by large scales and have a well-defined global structure: namely the perturbation is negative (decreasing $B_z$) on the edge of the layer and
positive (increasing $B_z$) in the middle of the layer, see Fig.~\ref{bz_slice}. Note that total $B_z$ flux must conserve. Why turbulence creates large-scale structure in $B_z$ so that it is larger in the center? This could be due to the diamagnetism of turbulence, which is stronger on the edge, where turbulence is more intense, so that the diamagnetism of turbulence has been pushing $B_z$ flux towards the center. This conjecture will require future research. The peculiar structure of the mean flux through the layer can have consequences for the particle acceleration in turbulent current layers as particles are more likely to be trapped in the low-$B_z$ regions.
When the mean field increases, the effect becomes negligible due to weaker coupling between
Alfv\'en and slow mode.

\section{2D vs 3D}
The results reported above appear to be qualitatively different from previous 2D results, which was expected:
the geometrical constraints in the 2D magnetic configuration naturally features magnetic separatrixes, X-points and magnetic islands, which are normally absent in 3D. Also, the dynamic influence of the global mean field, which is present in a generic reconnection geometry, is completely ignored by the 2D treatment. Another important point is that if $B_z$ tension plays no important role in 2D, the perturbations that govern 2D case are not Alfv\'enic and the arguments for the reduced MHD analogy I used in the above section are also not applicable.   

The 3D spontaneous reconnection that I studied here, proceeded in a different way than the 2D case, which, in \citet{loureiro2012} was dominated by the ejection of plasmoids and had significant time-dependence. In the 3D case, considered here, I observed a very steady rate (Fig.~\ref{width}) with turbulent current layer slowly eating through the mostly undisturbed fluid and turbulence being fueled by the free energy of the oppositely directed magnetic fields. 3D case was also different from 2D case in that the memory of the initial conditions, i.e. the location of the original current sheet was largely forgotten. In 2D case the current sheet remains precisely where it was, up to very high $S$, generating and ejecting plasmoids along the same line. In 3D case only small pieces of the original current sheet are visible after $t=10$ and the layer otherwise looks turbulent. Few large-scale structures in 3D may be called flux ropes, however, unlike 2D, they are turbulent inside (Fig.~\ref{b_slice}), also the number of these structures does not depend on $S$ as it does in 2D case described in \citet{uzdensky2010}. Another difference with 2D is the asymmetry of emerging turbulence
with respect to the original current sheet -- in 3D often the upper
or a lower part of the layer dominates. 

\begin{figure}[t]
\begin{center}
\includegraphics[width=0.8\columnwidth]{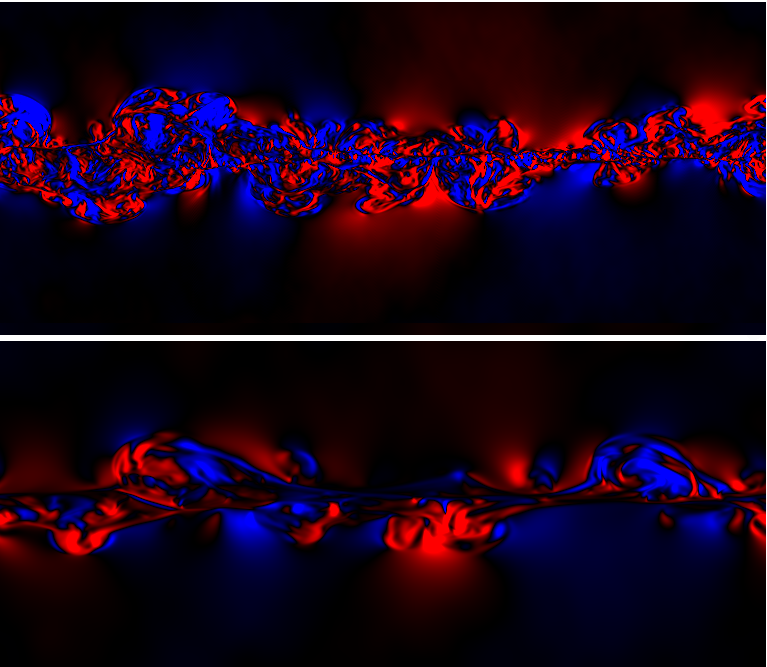}
\end{center}
\caption{The structure of $v_x$ in the x-y slice of the layer. The width of the picture is the box size. Red and blue show positive and negative $v_x$, for the same slices as on figure above.
The turbulent current layer generally lacks the bipolar inflow symmetry of the Sweet-Parker layer which is often observed between plasmoids in 2D.}
\label{vx_slice}
\end{figure}

The classic X-point inflow/outflow picture is usually preserved in 2D in each X-point between plasmoids.
In my simulations such a simple picture was not observed, see Fig.~\ref{vx_slice}.

The physical reason for resistivity-independent rate also appears to be different: in the 2D case it relies on the hierarchical formation and ejection of plasmoids \citep{uzdensky2010}, while in 3D I hypothesize this to be a consequence of turbulence locality, similar to models of reconnection due to ambient turbulence \citep{lazarian1999}. The difference between 3D reconnection with ambient turbulence and spontaneous
reconnection is that in the spontaneous case there is no external agent that drives reconnection
and there is no parameter of the amplitude of ambient turbulence as in, e.g., \citet{kowal2009}.

The resistively-independent turbulent reconnection has been already argued for reconnection due to ambient turbulence \citep{lazarian1999,eyink2011b}. In this paper I extend this result to the case where reconnection develops spontaneously without an external agent. I demonstrate this by showing that turbulence
in the current layer resemble ordinary turbulence and likely to be local in scale. At the
sufficiently large Lundquist numbers, the dynamics of large scales, which determines such properties as the bulk reconnection and dissipation rates, will be disconnected from the dynamics on plasma scales and dissipation parameters. It is, therefore, natural to expect reconnection and dissipation rates to go to asymptotic universal values. Whether the locality argument can be applied in 2D is unclear because large plasmoids may couple large and small scales directly.

The reconnection speed in ambient turbulence was argued to be proportional to
the kinetic energy density \citep{lazarian1999}. Our $v_r=0.015 v_{Ay}$ measurement obtained in absence of ambient turbulence could be seen as a lower limit
on fluid reconnection speed in 3D.

\section{ A model for the reconnection rate}
2D theory explaining observed reconnection rate is based on a hierarchical plasmoid model \citep{uzdensky2010}. This model predicts that the number of plasmoids scales linearly with $S$
and this hierarchy truncates on the scale of the critical layer. Thus the reconnection rate is just equal
to the SP rate at the critical value of $S$: $v_r=v_A S_{crit}^{-1/2}$. In my picture reconnection
is not due to large-scale tearing, but due to turbulence on the edge of the layer. This turbulence, as
I showed above is strong and can not be considered a linear stage of any instability. 

One of the interesting empirical facts about spontaneous reconnection that I observed in simulations
is that the reconnection rate is constant in time and the level of velocity and magnetic perturbations
keeps approximately on the same level as well. Can this be reconciled with the turbulent picture, despite the volumetric dissipation rate inside the layer depends on time?
The basic scaling for turbulent velocity as a function of turbulent cascade rate is
\begin{equation}
\delta v_l^2=C_{K,v} \epsilon^{2/3} l^{2/3},
\end{equation}
where $l$ is a scale of interest and I introduced Kolmogorov constant $C_{K,v}$ that refers to the
velocity perturbation, not the total energy spectrum. If we argue that turbulence is driven on the scale of the current layer thickness, i.e., $l=\Delta$, and use Eq.~\ref{eps_form} for the dissipation rate, we calculate that $\delta v_l^2$ indeed does not depend on $\Delta$ and, therefore, on time:
\begin{equation}
\delta v_l^2=C_{K,v} w_d^{2/3} v_r^{2/3} v_{Ay}^{4/3}.
\label{v_rms}
\end{equation}
The fact that $\delta v_l$ is constant in time is consistent with my numerical measurement.
How reconnection rate depends on $\delta v_l$ is not immediately obvious. One possibility is to use expression for turbulent reconnection rate from \citet{lazarian1999}, but it is not clear how
the layer width relates to the time-dependent injection scale $l$.
Assuming that $v_r$ depends only on the RMS velocity $\delta v_l$ at in the current layer, however, in the manner $v_r = v_{Ay} f(\delta v/v_{Ay})$ I can obtain time-independent rate by substituting Eq.~\ref{v_rms} and solving the nonlinear equation
\begin{equation}
v_r = v_{Ay} f(C_{K,v}^{1/2} w_d^{1/3} v_r^{1/3} v_{Ay}^{-1/3}).
\end{equation}
In particular, using $v_r \sim M_A^2$ dependence from \citet{lazarian1999}, e.g., choosing $v_r=C_{LV} \delta v_l^2/v_{Ay}$ I obtain
\begin{equation}
v_r=C_{LV}^3 C_{K,v}^3 w_d^2 v_{Ay}.
\label{rate_spon}
\end{equation}
The constant $C_{LV}$ has not been yet precisely measured \citep[c.f. ][]{kowal2009}. The constant
$C_{K,v}$ can be obtained in my simulations and refer to the ordinary Kolmogorov constant,
as well as the fraction of the total cascade energy that resides in its kinetic part. Introducing the ratio of kinetic to magnetic energy as $r_A$, which is around 0.13 in my simulations, I can estimate
$C_{K,v}=C_K r_A/(1+r_A)\approx 0.48$ using $C_K=4.2$ from \citet{B11}. This gives the reconnection rate of $0.018 C_{LV}^3 v_{Ay}$, which corresponds to the measured value if $C_{LV}=0.94$.

Interestingly, this expression depends only on the basic properties of well-developed turbulence, dimensionless numbers $C_K$ and $r_A$ and not on the properties of the instability.


\section{Electron acceleration}
The actual dissipation mechanisms of reconnection which will result in observable phenomena are still debated. It is plausible that dissipation in plasmas sometimes results in heating, and sometimes in the acceleration of fast particles. For example, the acceleration on shock fronts
starts with particles being pulled out of the thermal pool due to extremely high velocity gradient
at the shock front itself. Similarly, solar X-ray flares, which produce accelerated electrons
has been brought up as a proof that current layers must have microscopic widths to allow for plasma effects, including parallel electric field, and electron acceleration. The logic is the following: suppose that turbulent reconnection picture is true and current layers are wide compared with plasma scales and electrons only ``feel'' local turbulent perturbations, in which case
electron acceleration will be, basically, stochastic turbulent acceleration. Turbulent acceleration, specifically from the quasilinear theory \citep{Schlickeiser2002}, was calculated to be second order in $v/c$, too slow in many
practical cases, while acceleration by electric field $E \sim {\bf v_r \times B}/c$ is first
order in $v/c$ and should dominate. 

We recently found that the very basis of the above argument, the claim that turbulent acceleration must be
second order is, in fact, not true. In particular, we found analytically that if turbulence is
fueled by magnetic energy, like in the case of spontaneous turbulent reconnection, the structure of magnetic and electric fields in this turbulence is such that the average acceleration by curvature drift is positive. This is due to a mathematical relation between the MHD term which is responsible for energy transfer between kinetic and magnetic energy and the term responsible for the curvature drift acceleration \citep{BH16}. This does not require extra assumptions such as that particles have to be trapped for considerable time in magnetic islands, in fact the whole volume of turbulence
will be the first order accelerator. So, as long as the particle gyro-radius is smaller than the current layer width, the acceleration of these particles will be efficient.
The expression for the acceleration rate we derived in \citet{BH16},
\begin{equation}
 \frac{d{\cal E}}{dt}={\cal E_\|} \frac{8\pi}{B^2} \cal D,
\end{equation}
relates it to the energy transfer from magnetic to kinetic energy $\cal D$ and the acceleration rate. In ordinary driven turbulence this term is zero, while in spontaneous reconnection it is equal to the half of the total dissipation rate $\epsilon$. Since this mechanism results in average acceleration for all particles, all electrons are predicted to be accelerated to approximately the same energy, $0.35 T (L/d_i)^{1/4}$ \citep{BH16},
where $T$ is the thermal energy. The subsequent transition to the regime IV will result
in an outflow, particle escape, and additional acceleration due to converging magnetic field lines, which will result in a formation of a power-law tail.
It is interesting that X-ray emission during the solar flare indeed feature a thermal component 
and a power-law tail.

\section{Universal fluid resistance to thin current in the limit of zero resistivity}
Our result suggests that in the high-$S$ limit all macroscopic properties of reconnection are expressed in terms of macroscopic plasma properties $\rho$, $v_A$ and $L$
and independent of microscopic dissipation. This result is quite spectacular considering that individual field lines reconnection does depend on microphysics. 

Let us think of the turbulent current layer as a conductor. The electromotive force (EMF) between
two points separated by a large distance in $z$ direction will be expressed as
\begin{equation}
 {\cal E}=\int \frac{\bf v \times B}{c}\, dz \approx \frac{v_r}{c} B_{y0} L_z,
\end{equation}
and should be independent of whether we took point inside the current sheet or outside of it.
At the same time, the total current flowing through the width $L_y$ of the current layer
will be $I=(c/4\pi) 2 B_{y0} L_y$. Taking the ratio of the two, we obtain the effective
resistance of the current layer
\begin{equation}
 R =\frac12\frac{v_r}{c}\frac{L_z}{L_y}\left(\frac{4\pi}{c}\right) \approx 0.0075 \frac{v_A}{c}\frac{L_z}{L_y}\left(\frac{4\pi}{c}\right)
\end{equation}
Note that $R_0=4\pi/c$ ($\sim 376.73 \Omega$ in SI units) is known as the impedance of free space, 
and our final result is independent of microscopic resistivity. Such a
resistance will dissipate energy in conductive fluids in spite of the fact that microscopic resistivity of plasma in most astrophysical objects can be considered negligible.
In the limit of infinitely heavy plasma, $v_A/c \to 0$, this will result in zero resistance,
consistent with ordinary resistance expression proportional to resistivity. For the
very light plasma, i.e. the relativistic force-free magnetically dominated
limit $v_A/c=1$ and the resistance is a sizable fraction of the impedance of free space. 

For example, jets in active galactic nuclei are self-contained electromagnetic structures carrying large-scale poloidal current, with at least part of the return poloidal current flowing in a layer separating magnetic pressure-dominated jet and the outside medium \citep{begelman1984}. Poynting-dominated jet has an impedance of $\sim 90 \Omega$ \citep{lovelace2013},
and since $v_A/c\sim 1$ for rarefied electron-positron plasma, we can estimate that
jets with aspect ratios of $L_z/L_y > 650$ will dissipate a sizable
fraction of their energy in a outer current layer, due to this layer's fundamental fluid resistance.
Whether this will result in an outer layer's visibility is an open question, but considering
the result of the previous section, the first order acceleration of particles and non-thermal
emission from this layer is highly likely. 

Another example is dissipation in pulsar magnetospheres, which feature the return current layer in the equatorial plane beyond the light cylinder \cite{uzdensky2012}. The current layer separating open and closed field lines within light cylinder \citep[see, e.g.,][]{arons2011} can also result in acceleration.

\section{Discussion}
Our simulations clearly demonstrate that turbulence must be a part of high-Lundquist reconnection, e.g., astrophysical reconnection.
Quite importantly, this turbulence is not random, but contains non-trivial correlations
that comes from the fact that energy is transferred from magnetic to kinetic, these correlations
likely to result the efficient particle acceleration \citep{BH16}, which can help explaining why reconnection on the Sun results in powerful X-ray flares.
Observational evidence favoring spontaneous turbulent reconnection include magnetospheric observations that showed an enhanced level of turbulence inside current sheets \citep{matsumoto2003,cattell2005}.

Many astrophysical objects, such as the interstellar medium in our Galaxy, feature relatively high level of ambient turbulence and the reconnection is argued to be fast \citep{lazarian1999} due to the existing magnetic field stochasticity. In highly magnetized environments, such the solar surface or the pulsar wind nebulae, the velocity of ambient turbulence may be tiny compared to the local Alfv\'en speed and the turbulence, spontaneously generated by the current sheet, and fueled by the reconnecting field itself
is more important. More qualitatively, if $M_A= \delta v/v_A < \sqrt{0.015} \approx 0.12 $, spontaneous reconnection will dominate. Future parameter study in simulations 
with Lundquist number above critical and varying level of ambient turbulence should clarify the
transition between turbulent reconnection due to ambient turbulence and spontaneous reconnection.
So far, simulations with driven turbulence had $S<10^4$, so the non-driven case was still consistent with Sweet-Parker picture \citep{kowal2009}, with resistive rate higher than $0.015$ that I measured
in this paper.

Currently, two completely opposite ways to explain fast spontaneous reconnection exist. First is the model that relates reconnection rate to the critical Lundquist number, i.e. the claim that reconnection rate depends on the stability to linear resistive tearing. This gives dimensionless rate of $S_{crit}^{-1/2}$ \citep{uzdensky2010}. The second relates reconnection to the inherent properties of strong turbulence and gives the reconnection rate of $C_K^3 r_A^3/(1+r_A)^3$ (this paper). Both pictures reasonably agree with the measurement, but hard to reconcile with each other.
One way to connect these pictures is to imagine that turbulence produces constant anomalous resistivity which brings effective Lundquist number to or below critical value, just to barely suppress
tearing. The counter-argument to this is that the linear stability studies have
only been performed in the laminar SP regime and the large value of $S_{crit}$ is the result
of a rather non-trivial interplay between resistive tearing rate $\sim \eta^{1/2}$, and
thinning of the current layer. In a non-laminar case it is easy to argue for $S_{crit}\sim 1$, but hard to argue for $S_{crit}\sim 10^4$.


In stationary reconnection with speed $v_r$ that developed outflow with speed $v_{\rm out}$ and reach quasi-stationary state as in Fig.~\ref{layer} panel (IV) the inflow speed is balanced by the outflow $v_{\rm in}=v_r$. In this case the current layer width also reach stationary value of
$\Delta=L v_r/v_{\rm out}$, neglecting compressibility. Similar to the cascade locality argument, e.g.,
the argument why $v_r$ should be independent on $S$, the outflow speed also must be a fraction
of $v_A$. It is not necessarily equals to $v_A$ as it is often assumed. The outflow develops under the
force of magnetic tension from $B_y$ component but we showed that a sizable fraction of this energy
is dissipated and not converted into kinetic motion. We expect the outflow speed to be a sizable
fraction of the $v_{Ay}$, this fraction being lower by a factor of $\sim\sqrt{1-0.4}\approx 0.77$, where $w_d\approx 0.4$ is a dissipation factor that we found in this paper. 

The reconnection rate could be affected by the presence of an outflow in regime IV.
Increasing the box size will allow larger-scale fluid motions which could emulate the outflow effect locally, this also correspond to higher $S$. We showed that with increasing $S$ the reconnection rate goes to a constant. This is probably associated with the fact that most of the activity which results in a growth of the current layer happens on the boundary. Likewise, the simulation naturally included the
effects of the local outflows that develop on larger and larger scales as reconnection progresses, while
the reconnection rate stay relatively stable (Fig.~3).

It is also worth mentioning that the fraction of the dissipated energy that we measured, $w_d\approx 0.4$, will limit the compression ratio of the current layer. Previously it was thought that most of the magnetic energy during reconnection could be spent to accelerate the outflow jet, which is why the
outflow speed was always estimated being equal to Alfv\'en speed (see above). In this case the compression ratio may be arbitrarily high for low beta plasmas, going as $1/\beta$. In our case
the compression ratio will be limited to $1/w_t(\gamma-1)\approx 3.8$ for monoatomic gas.


{\bf Acknowledgements} I am grateful to Alex Schekochihin, Nuno Loureiro, Fan Guo, Alex Lazarian, Ethan Vishniac,
Homa Karimabadi, Bill Daughton and Hui Li for fruitful discussions. Computations were performed on NICS Kraken with XSEDE
allocation TG-AST110057 and on LANL institutional computing resources.

\def\araa{{\rm Ann. Rep. A\&A} }
\def\apj{{\rm ApJ}}           
\def\apjl{{\rm ApJ}}          
\def\apjs{{\rm ApJ}}          
\def\grl{{\rm GRL}}
\def\aap{{\rm A\&A}}
\def\mnras{{\rm MNRAS}}
\def\physrep{{\rm Phys. Rep.}}               
\def\prl{{\rm Phys. Rev. Lett.}} 
\def\pre{{\rm Phys. Rev. E}} 
\def\nat{{\rm Nature}} 
\def\jgr{{\rm JGR}}
\bibliography{all}

\end{document}